\documentstyle[12pt,graphicx]{article}

\begin{document}

\title{Fermionic microstates within
Painlev\'e-Gullstrand black hole.}
\author{P. Huhtala$^{1}$ and G.E. Volovik$^{1,2}$
\\$^{1}$ Low Temperature Laboratory, Helsinki University of
Technology\\
P.O.Box 2200, FIN-02015 HUT, Finland\\
\\
$^{2}$ L.D. Landau Institute for
Theoretical Physics\\  Kosygin Str. 2, 117940 Moscow, Russia
}
\maketitle
\begin{abstract}
{We consider the quantum vacuum of fermionic
field in the presence of a black-hole background as a possible candidate
for the stabilized black hole. The stable vacuum state (as
well as thermal equilibrium states with arbitrary temperature) can exist
if we use the Painlev\'e-Gullstrand description of the black hole, and the
superluminal dispersion of the particle spectrum at high energy, which is
introduced in the free-falling frame. Such choice is inspired by the
analogy between the quantum vacuum and the ground state of quantum
liquid, in which the event horizon for the low-energy fermionic
quasiparticles also can arise.  The quantum vacuum is characterized by
the Fermi surface, which appears behind the event horizon. We do not
consider the back reaction, and thus there is no guarantee that the
stable black hole exists. But if it does exist, the Fermi surface
behind the horizon would be the necessary attribute of its vacuum state.
We also consider exact discrete spectrum of fermions inside the horizon
which allows us to discuss the problem of fermion zero modes.
 }
\end{abstract}

\section{Introduction}

In 1981 Unruh suggested to study the black hole physics
using its sonic analogue \cite{UnruhSonic}. Originally suggested for
classical liquids, this was later extended to quantum systems such as
superfluids and Bose condensates
\cite{JacobsonVolovik,Garay,ChaplineLaughlin}.
The main advantage of the quantum liquids and gases is that in many
respect they are similar to the quantum vacuum of fermionic and bosonic
fields. This analogy forms a view on the quantum vacuum as a special type
of condensed matter -- the `ether' --  where the physical laws which we
have now can arise emergently as the energy or temperature of the `ether'
decreases
\cite{LaughlinPines}.  The particular scenario of the emergent formation
of the effective gravity together with gauge fields and chiral fermions
can be found in the recent review paper
\cite{Volovikreview}.

According to the topology in the momentum space, there are
three types (universality classes) of the fermionic vacua. One of them
has trivial topology and as a result its fermionic excitations  are fully
gapped (massive fermions). The other two have nontrivial momentum-space
topology characterized by certain topological invariants in the momentum
space \cite{Volovikreview}. One of the  two nontrivial
universality classes contains systems with Fermi Points; their excitations
are chiral fermions, whose energy turns to zero at points in the momentum
space. Another class represents systems with wider manifold of
zeroes: their gapless fermionic excitations
are concentrated in the vicinity of the 2D surface in momentum space --
the Fermi Surface. This class contains Fermi liquids.

Here we discuss the properties of the quantum vacuum in the presence of
event horizon.  We assume that in the absence of horizon the
fermionic vacuum  belongs either to the trivial class (such as the
Standard Model  below the electroweak transition where all fermions are
massive) or to  the class of Fermi Points (such as the Standard Model
above the electroweak transition whose excitations are chiral massless
fermions).

In the presence of a horizon the region behind the horizon becomes the
ergoregion: the particles acquire negative energy there. In the true
vacuum state these negative energy levels must be occupied, which means
that the old vacuum must be reconstruced by filling these levels. We do
not study the process of the filling -- it can be the smooth process of
Hawking radiation \cite{HawkingNature} or some other more violent process
-- we discuss what will be the structure of the true vacuum state if it is
possible to reach this state without destruction of a horizon. In other
words we assume that the stable black hole can exist as a final ground
state of the gravitational collapse. We find that behind
the horizon the fermionic vacuum belongs to the class of the Fermi
Surface.

The main sources for the appearance of the Fermi Surface come from the
following properties of event horizon. First,  the emergence of the
Planck physics in the vicinity of (and behind) the horizon. Event horizon
serves as a magnifying glass through which the phenomena at Planck length
scale could be visualized.  At some scales the Lorentz invariance -- the
property of the low-energy physics -- inevitably becomes invalid and
deviations from the linear (relativistic) spectrum become important. Such
violation of Lorentz invariance is now popular in the literature
\cite{UnruhSonic,BHdispers,Corley,BHlaser,Starobinsky,JacobsonDispersion}.
It leads to either subluminal or superluminal propagation at high
energy, say, $E^2(p)=c^2p^2(1 \pm  p^2/p_P^2)$, where $p_P$ is Planck
momentum.   According to the condensed matter analogy we assume that the
high energy (quasi)particles are superluminal, i.e. the sign is plus.
Due to the  superluminal dispersion there is a bottom in the Dirac sea
and thus the process of the filling of the negative energy levels becomes
limited.  When all of them are occupied, we come to a global vacuum state
(or global thermodynamics equilibrium with positive heat capacity, if the
temperature is finite). Thus the superluminal dispersion of the particle
energy gives rise to the energetic stability of the vacuum in the
presence of black hole.

The second important consequence of the event horizon, due to which
the vacuum belongs to the class of systems with the Fermi
Surface, is that the horizon violates the time reversal symmetry of the
system: the ingoing and outgoing particles have different trajectories. In
condensed matter the appearance of the Fermi surface due to violation of
the time reversal symmetry is a typical phenomenon (see e.g.
\cite{FermiSurfaceSuperconductor} and also Sec. 12.4 of Ref.
\cite{Volovikreview}).

In Refs.
\cite{ChaplineLaughlin,MazurMottola} the stable black hole is also
considered, which exhibits a finite positive heat capacity, any
temperature it likes, and no Hawking radiation. But it is assumed
there that in the final state the time reversal symmetry is not broken (or
actually it is restored in the final state). The existence of such stable
black hole with unbroken time reversal symmetry is also supported by the
condensed matter analogies \cite{ChaplineLaughlin,Mohazzab,VierbeinWalls}
in which stable infinite-redshift surface arise. Example of the
infinite-redshift surface with no time reversal symmetry breaking is also
provided by the extremal black hole, whose condensed matter analog is
discussed in Sec. 12.6 of review
\cite{Volovikreview}. In all these examples the
Fermi surface does not appear. The black hole ground states with time
reversal symmetry are in some sense exceptional (in the same manner as
extremal black hole) and we shall not discuss them here.

\section{Stationary metric with explicit violation of time reversal
symmetry}

The vacuum can be well defined only if the metric is stationary. In
general relativity the stationary metric for the black hole is provided
in the Painlev\'e-Gullstrand spacetime \cite{Painleve}. The line
element of the Painlev\'e-Gullstrand metric is
\begin{equation}
 ds^2=- c^2dt^2+ (d{\bf r}-{\bf v}dt)^2 = - (c^2-
v^2)dt^2- 2{\bf v}d{\bf r}dt+d{\bf r}^2~,
\label{Painleve}
\end{equation}
where
\begin{equation}
{\bf v}({\bf r})=\pm \hat{\bf r}c\sqrt{r_h\over r} ~,~r_h={2MG\over c^2}~.
\label{VelocityField}
\end{equation}
Here $M$ is the mass of the hole; $r_h$ is the radius of the horizon;
 $G$ is the Newton
gravitational constant; the minus sign in
Eq.(\ref{VelocityField}) gives the metric for the black hole; while
the plus sign characterizes the white hole. The time reversal operation
$t\rightarrow -t$ transforms the black hole into white whole.
The stationarity of this metric and the fact that it describes the
spacetime both in exterior and interior regions, are very attractive
features and they became explored starting from the Ref.
\cite{KrausWilczek94} (see
\cite{Martel,Schuetzhold,ParikhWilczek}; extension of
Painlev\'e-Gullstrand spacetime to rotating black hole can
be found in Ref. \cite{Doran}).

In case of the black hole the field ${\bf v}({\bf r})$ has simple
interpretation: it is the velocity of the observer who freely falls along
the radius towards the center of the black hole with zero initial
velocity at infinity. The motion of the observer obeys the Newtonian laws
all the way through the horizon
\begin{equation}
{d^2r\over dt^2}=-{GM\over r^2}~,
\label{NewtonianLaws1}
\end{equation}
and thus his velocity is
\begin{equation}
{\bf v}({\bf r})\equiv {d{\bf r}\over dt}=-\hat{\bf r}\sqrt{2GM\over
r} ~.
\label{NewtonianLaws2}
\end{equation}
The time coordinate $t$ is the
local proper time for the observer who drags the inertial coordinate frame
with him.

As was first noticed by Unruh
\cite{UnruhSonic}, the effective metric of the type in Eq.(\ref{Painleve})
is experienced by quasiparticles propagating in moving fluids. The field
${\bf v}({\bf r})$ is just the velocity field of the liquid, and $c$ is
the `maximum attainable velocity' of quasiparticles in the low-energy
limit, for example the speed of sound in case of phonons (see also
\cite{Visser,Visser2,Stone,Volovikreview,SakagamiOhashi}). The horizon
 could be produced in liquids when the
flow velocity becomes bigger than $c$.
The black hole and the white hole can be reproduced by the liquid flowing
radially inward and outward correspondingly. This is an explicit
realization of the breaking of the time reversal symmetry by flowing
liquid: time reversal operation reverses the direction of flow of the
`vacuum': ${\cal T}{\bf v}({\bf r})=-{\bf v}({\bf r})$.

This  Painlev\'e-Gullstrand spacetime, though not static, is stationary.
That is why the energy $\tilde E$ of (quasi)particle in this spacetime is
determined both in exterior and interior regions.  It can be obtained as
the solution of equation
$g^{\mu\nu}p_\mu p_\nu + m^2=0$ with $p_0=-\tilde E$, which gives
\begin{equation}
\tilde E({\bf p})=  E(p)   +{\bf p}\cdot {\bf
v}({\bf r}) ~,
\label{EnergySpectrum1}
\end{equation}
where  $E(p)$ is the energy of the particle in the free-falling
frame:
\begin{equation}
E^2(p)= p^2c^2 +m^2~.
\label{EnergySpectrum2}
\end{equation}
 For the `sonic' black hole it is
the energy of the quasiparticle in the frame comoving with the superfluid
vacuum.

Let us consider a massless (quasi)particle moving in the radial direction
from the black hole horizon to infinity, i.e. its radial momentum $p_r>0$.
Since the metric is stationary the energy of a particle in the
Painlev\'e-Gullstrand frame (or of a  quasiparticle in the laboratory
frame) is conserved and one has $\tilde E=Const$. Then its energy in the
free-falling (superfluid comoving) frame:
\begin{equation}
E(p) =cp_r= {\tilde E\over 1+  v(r)/c} = {\tilde E\over 1
-\sqrt{r_{\rm h}\over r}}~.
\label{BlueShiftEnergy}
\end{equation}
This energy, which is very big
close to the horizon, becomes less and less when the (quasi)particle moves
away from the horizon. This is the gravitational
red shift superimposed on the Doppler effect
\cite{LandauLifshitz2}, since the emitter is free falling with the
velocity $v=v_{\rm s}(r)$. The frequency of the spectral line  measured
by the observer at infinity is
\begin{equation}
\tilde\omega=\omega\sqrt{-g_{00}}~{\sqrt{1-{v^2\over c^2}}\over
1-{v\over c}}=\omega\left(1 -\sqrt{r_{\rm h}\over r}\right)~,
\label{RedShiftFrequency}
\end{equation}
 where $\omega$ is the nominal frequency of
this line. The surface $r=r_{\rm h}$ is a surface of infinite redshift, at
this surface the energy in Eq.(\ref{BlueShiftEnergy}) diverges.  This
means that if we observe particles coming to us from the very vicinity of the
horizon, these outgoing particles originally had a huge energy approaching the
Planck energy scale. Thus the event horizon can serve as a magnifying glass
which allows us to see what happens at the Planck length scale.  At some
point the low-energy relativistic approximation inevitably becomes invalid
and the Lorentz  invariance is violated.

In quantum liquids the nonlinear dispersion enters the velocity
independent energy
$E(p)$ in the superfluid comoving frame. Taking  into account the analogy
with quantum liquids, we assume that in our vacuum the Planck physics also
enters the energy in the free-falling frame. Thus the energy spectrum of
the particles is given by Eq.(\ref{EnergySpectrum1}) where
\begin{equation}
 E^2(p)=m^2+ p^2c^2 \left(1 \pm  {p^2\over p_P^2}\right)~.
\label{EnergySpectrumNonlinear}
\end{equation}

As for the ingoing particle, its radial momentum $p_r<0$ and thus its
energy in the comoving frame
\begin{equation}
E(p) =-cp_r= {\tilde E\over 1-  v(r)/c} = {\tilde E\over 1
+\sqrt{r_{\rm h}\over r}}~.
\label{BlueShiftEnergyIngoing}
\end{equation}
It has no pathology at the horizon -- the observer falling
freely across the horizon sees no inconveniences when he crosses the horizon
-- and thus the Planck physics is not envoked here.

The pathology reappears when one tries to construct  the thermal
state of global equilibrium (or the vacuum state) in the presence of
a horizon. According to the Tolman law in the global equilibrium the
temperature as measured by observer in the comoving frame diverges at the
horizon:
\begin{equation}
T_({\bf r})={T_{Tolman}\over
\sqrt{-g_{00}({\bf r})}}={T_{Tolman}\over \sqrt{ 1 - {v^2 \over c^2}}}\,.
\label{TolmanLaw}
\end{equation}
Again at some point this temperature becomes so high that the Planck
physics becomes relevant. The global equilibrium in the presence of a
horizon is possible only for the superluminal dispersion, i.e. for the
sign plus in Eq.(\ref{EnergySpectrumNonlinear}).  The reason is the
following.  Behind the horizon, at $r<r_{\rm h}$, the velocity of the
frame-dragging exceeds the speed of light. In the relativistic domain
this means that the radial coordinate $r$ becomes time-like, because a
(quasi)particle behind the horizon can move along the $r$ coordinate only
in one direction: towards the singularity. However, with the plus sign
for the energy spectrum in Eq.(\ref{EnergySpectrumNonlinear}) the
(quasi)particles can go back and forth even behind the horizon. Thus the
spacelike nature of the $r$-coordinate is restored by the superluminal
dispersion, and the global equilibrium becomes possible.

Finally, the condensed matter analog of formation of quantum field theory
as emergent phenomenon at low-energy, suggests that our vacuum is
fermionic, while all the bosonic degrees of freedom can obtained as
collective modes of the fermionic vacuum. It is the Pauli principle for
fermions, which allows us to construct the stable vacuum in the
preence of a horizon. Thus there are 3 main necessary conditions for the
existence of stable vacuum with broken time reversal symmetry in the
presence of black hole:  vacuum is fermionic, its fermionic
excitations have superluminal dispersion, the black hole is described by
the Painlev\'e-Gullstrand metric.  All three conditions are  motivated by
the quantum liquid analogies.

\section{Dirac equation in Painlev\'e-Gullstrand metric}

In Ref. \cite{Volovik} fermions have been considered in
semiclassical approximation. Here we extend this consideration to the
exact quantum mechanical. Fermions in the presence of the nontrivial
gravitationa background are described by the tetrad formalism. We  follow
here Ref.
\cite{Weinberggravcos}. The metric $g_{\mu
\nu}$ can be written in terms of   tetrads $e^{a}_{\mu }$:
\begin{eqnarray} g_{\mu
\nu}=e^{a}_{\mu } e^{b}_{\nu }
\eta_{a b} \end{eqnarray}
where $\eta^{ab}=diag(-1,1,1,1)$.
The Dirac equation in curved spacetime is
\begin{equation}
\left(i\gamma^aE^{\mu}_aD_\mu
-m\right)\Psi=0~,~D_\mu=\partial_\mu +{1\over
4}\omega_{\mu;ab} \gamma^a\gamma^b~,
\label{DiracWithTorsionField}
\end{equation}
where the
dual tetrad field $E^{\nu}_b$ obeys:
\begin{eqnarray}
\label{uusi1} && g_{\mu\nu}= e^a_\mu e^b_\nu
\eta_{ab}~,~E^{\mu}_ae^a_\alpha=\delta^\mu_\nu~,
~E^{\mu}_aE^{\nu}_b\eta^{ab}=g^{\mu\nu}~,\\ \label{uusi2} &&e^a_\mu=
g_{\mu\nu}\eta^{ab}E^\nu_b~,~e_{\nu
b}=e_{\nu}^a\eta_{ab}=g_{\mu\nu}E^{\mu}_b~,
\end{eqnarray}
and the torsion field is
 \begin{equation}
\omega_{\mu;ab}=E^{\nu}_a\eta_{bc}\nabla_\mu e^c_\nu=
E^{\nu}_a\nabla_\mu\left(g_{\nu\alpha}E^{\alpha}_b\right)=E^{\nu}_a
\nabla_\mu e_{\nu b}= E^{\nu}_a\left( \partial_\mu e_{\nu
b}-\Gamma^{\gamma}_{\mu\nu} e_{\gamma b}\right) ~.
\label{TorsionField}
\end{equation}
The vielbeins which correspond to the general `flow' metric  in
Eq.(\ref{Painleve}) are
\begin{equation}
 e^a_\mu=\delta^a_\mu +\tilde  e^a_\mu~,~\tilde
e^a_\mu= v^{i}\delta^a_i\delta^0_\mu ~.
\label{TetradsDecart}
\end{equation}
The only nonzero correction to the tetrad field
$\delta^a_\mu$ for Minkowski spacetime is
$\tilde e^i_0=v^{i}\neq 0$.
For the Painlev\'e-Gullstrand metric of the black hole in spherical
coordinate system one has
\begin{equation}
 e^0_\mu=(1,0,0,0)~,~ e^1_\mu=(v,1,0,0)~,~
e^2_\mu=(0,0,r,0)~,~ e^3_\mu=(0,0,0,r\sin\theta)~,
\label{DoranEq27tetrads}
\end{equation}
where $v(r)=-r^{-1/2}$, assuming $c=r_h=1$.

The violation of the Lorentz invariance at high
energy can be introduced by adding the nonlinear
 $\gamma_5$-term, which gives the superluminal dipersion. As a result one
has the  Dirac equation in the Painlev\'e-Gullstrand metric \cite{Doran},
which is now modified by the non-Lorentzian term:
\begin{eqnarray}
 i\partial_t \Psi = -ic\alpha^i\partial_i \Psi +m \gamma_0 \Psi
 +H_P \Psi+H_G \Psi ~.\end{eqnarray}
Here $H_P$ and $H_G$ are Hamiltoinians coming from the Planck physics
and  from the gravitational field  correspondigly:
\begin{eqnarray}
H_P=-{c\over p_P} \gamma_5 \partial_i^2 ~,~
H_G   =ic \sqrt{{r_h \over r}} \left( {3\over 4r}
  + \partial_r   \right) ~.
\label{HPHG}
\end{eqnarray}
The $\gamma$ matrices used are
\begin{eqnarray} \alpha^i= \left(
\begin{array}{cc} 0&\sigma^i \\ \sigma^i & 0 \end{array} \right) \ \ \
\   \   \  \gamma^0= \left( \begin{array}{cc} 1 & 0 \\ 0 & -1 \end{array}
\right) \end{eqnarray}
and
\begin{eqnarray} \gamma_5=i\gamma_0 \gamma_1 \gamma_2 \gamma_3= \left(
\begin{array}{cc} 0&-i \\ i& 0 \end{array} \right) \end{eqnarray}
After
multiplication by $r_h/(\hbar c)$ we get a dimensionless form and write
$\hbar =c=r_h=1$ and $p_0= p_Pr_h/\hbar \gg 1$.

\section{Eigen states of fermions in Painlev\'e-Gullstrand black hole}

Since $\partial_t$ is the timelike Killing vector in the
Painlev\'e-Gullstrand black hole the energy
$\tilde E$ is well defined quantity, and  the variables
$t$ and
${\bf r}$ can be separated by writing
\begin{eqnarray}
\label{eka1.5}
\Psi=\left( \begin{array}{c} \phi({\bf r}) \\ \chi(
{\bf r} ) \end{array} \right) e^{-i\tilde Et} ~,
\end{eqnarray}
 The ${\bf r}$-equations are now
\begin{eqnarray}
\nonumber
\tilde E\phi= {\bf \sigma } \cdot {\bf p} \chi
+m\phi-i{1\over p_0}p^2 \chi+ H_G\phi \\
\nonumber \tilde E\chi ={\bf \sigma }
\cdot {\bf p} \phi -m\chi +i{1\over p_0} p^2 \phi +H_G\chi ~,\\
\end{eqnarray}
where $p_i=-i\partial_i$.
Using the spherical symmetry one introduces in the standard way the
spherical harmonics -- eigenstates of the operators ${\bf J}^2$
and $J_z$ -- where ${\bf J}$  is the total angular momentum
 \begin{eqnarray} J_i=L_i+S_i=L_i+{1\over 2} \left(
\begin{array}{cc} \sigma_i & 0 \\ 0 & \sigma_i \end{array} \right)
\end{eqnarray}
and $L_i$ is the orbital angular momentum operator in $R^3$.
Since we will be interested in the states with high momenta $J \sim p_0
\gg mr_h/\hbar$  we can neglect the mass term. Then one obtains the
following ansatz:
\begin{eqnarray}
\label{eka1.18}
\phi_{J,J_3}={1\over 2r}\left( (f^+(r)+f^-(r))\Omega_{l} +
(f^+(r)-f^-(r))\Omega_{l+1} \right)~,\\
\label{eka1.19}
\chi_{J,J_3}={1\over 2r}\left( (g^+(r)-f^+(r))\Omega_{l} +
(g^+(r)+g^-(r))\Omega_{l+1} \right)~.
\end{eqnarray}
Here the spherical harmonics are
\begin{eqnarray}
 \Omega_l =
\left( \begin{array} {c}
\sqrt{ {J+J_z \over 2J} } Y_{l,J_z-1/2} \\ \sqrt{ {J-J_3
\over 2J} } Y_{l,J_3+1/2} \end{array} \right) ~,~   \Omega_{l+1} =
\left( \begin{array} {c} -\sqrt{ {J-J_3 +1 \over 2J+2} } Y_{l+1,J_3-1/2}
\\
\sqrt{ {J+J_3+1 \over 2J+2 } } Y_{l+1,J_3+1/2} \end{array} \right)~,
\end{eqnarray}
where $l=J-1/2$. The radial functions satisfy the following equations:
 \begin{eqnarray}
\nonumber
 &&\tilde E\left( \begin{array} {cc} f^+
\\ g^+ \end{array} \right) = \left[ i\partial_r \left( \begin{array} {cc}
0&1 \\ 1&0 \end{array} \right) +i{l+1\over r} \left( \begin{array}
{cc} 0 & 1 \\ -1 & 0\end{array} \right) + {l+1\over p_0 r^2}+\right. \\
\nonumber && \left.   + {1\over p_0} \left( -\partial_r^2+{(l+1)^2
\over r^2} \right) \left( \begin{array} {cc} -1 & 0 \\ 0 & 1
\end{array} \right) +i\sqrt{1/r} \left( \partial_r-{1\over 4r} \right)
\right] \left( \begin{array} {cc} f^+ \\ g^+ \end{array} \right) \\
\label{toka1.44}  \\
\nonumber &&\tilde  E\left( \begin{array} {cc} f^- \\ g^- \end{array}
\right) =\left[ i\partial_r \left( \begin{array} {cc} 0&1 \\ 1&0
\end{array}
\right) +i{l+1\over r} \left( \begin{array} {cc} 0 & 1 \\ -1 & 0
\end{array} \right) - {l+1\over p_0 r^2}+ \right. \\ \nonumber
&& \left. + {1\over p_0} \left( -\partial^2_r+ {(l+1)^2\over r^2}
\right) \left( \begin{array} {cc} 1&0 \\ 0&-1
\end{array} \right)  +i\sqrt{1/r} \left( \partial_r-{1\over 4r} \right)
\right] \left( \begin{array} {cc} f^- \\ g^- \end{array} \right)
\\ \label{toka1.45} \end{eqnarray}
By complex conjugation one gets from (\ref{toka1.44}) the equation
(\ref{toka1.45}) with a change in the sign of energy.
This means that
the matrices cannot be diagonalised simultaneously, unless $\tilde
E=0$, thus the eigen state with $\tilde
E\neq 0$ has either nonzero $(f^+,g^+)$ or nonzero
$(f^-,g^-)$.

These equations (\ref{toka1.44}) and (\ref{toka1.45}) are starting
point for our consideration of the fermionic vacuum and excitations.

\section{\label{bssec} Fermions in semiclassical approximation}

In the classical limit, when $(f,g)\propto \exp{(i\int p_rdr)}$, one
obtains the following energy spectrum:
\begin{equation}
\left(\tilde E  +{p_r\over \sqrt{r}}\right)^2 = p_r^2+ {l^2\over r^2} +
{1\over p_0^2}\left( p_r^2+ {l^2\over r^2}\right)^2~.
\label{classicalEq}
\end{equation}
Here we neglected small terms
of the relative order $1/p_0$. We are interested in the states with the
lowest energy, since they give the main contribution to the
thermodynamics. For given
$l$ the energy of the fermion becomes zero at the following values of the
radial momentum:
\begin{equation}
  p_r^2(r,\tilde E=0,l)= {1\over 2r}p_0^2 (1-r) -{l^2\over
r^2}
\pm {1\over r}\sqrt{{1\over 4}p_0^4 (1-r)^2 -{p_0^2  l^2\over r}}~.
\label{SuperluminalTrajectoryZeroE}
\end{equation}
This coincides with Eq.(13) of Ref. \cite{Volovik}, where the
quasiclassical approximation has been used from the very beginning.

\begin{figure}
  \centerline{\includegraphics[width=0.9\linewidth]{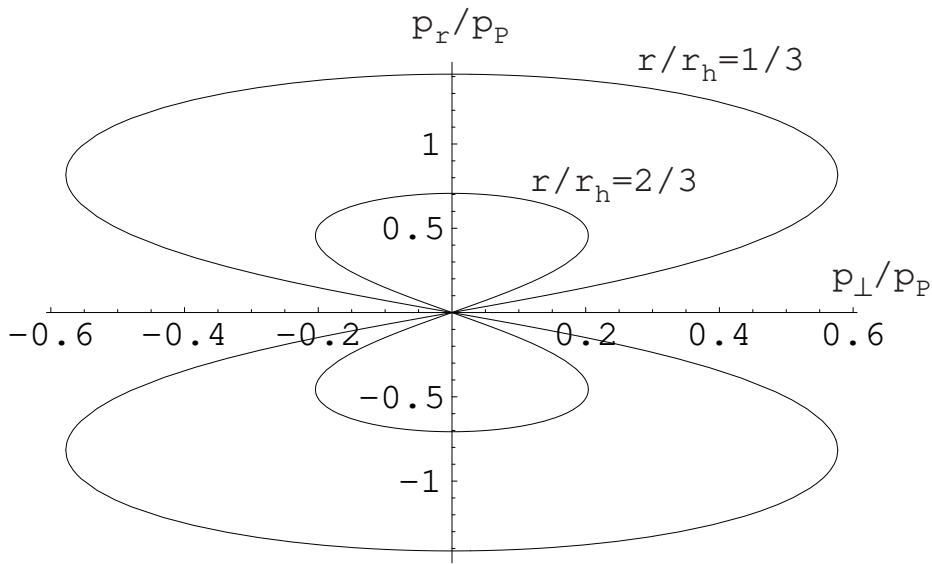}}
  \caption{Fermi surface $\tilde E({\bf p})=0$ at two positions inside the
black hole: $r=2r_h/3$ and $r=r_h/3$. }
  \label{FermiSurfaceFig}
\end{figure}

In the completely classical consideration, when $p_\perp=l/r$ represents
the transverse momentum of the fermion, the Eq.(\ref{classicalEq}) at
$\tilde E=0$ gives the closed 2D surface in the 3D momentum space. This
surface where the energy of particles is zero represents the Fermi
surface, it exists only inside the horizon, i.e. at
$r<r_h$ ($r<1$).  Fig. \ref{FermiSurfaceFig} demonstrates the Fermi
surface
$\tilde E({\bf p})=0$ at two values of the radius $r$ behind
the horizon:
$r=2r_h/3$ and $r=r_h/3$. The area of Fermi surface increases with
decreasing
$r$.

In the true ground state all the levels inside the Fermi surface,
i.e. with $\tilde E({\bf p})<0$, must be occupied. Of course, such
reconstruction of the vacuum, which involves the Planck energy scale,
can have tremendous consequence for the black hole itself. This
cannot be described by the phenomenological low-energy physics.
Nevertheless, we can claim that if the horizon still survives after
the vacuum reconstruction, the Fermi surface will also survive because of
its topological robustness. In this case the statistical physics of the
black hole microstates will be completely determined by the fermionic
states in the vicinity of the Fermi surface. In particular, the entropy
and the heat capacity of the black hole are linear in temperature $T$:
\begin{equation}
  S=C={\pi^2\over 3}N(0)T~,
\label{EntropySpecificHeat}
\end{equation}
where $N(0)$ is the density of states at $\tilde E=0$. From the general
dimensionality arguments together with the fact that the density of
states must be proportional to the volume of the Fermi liquid one
obtains
\begin{equation}
N(0)=\gamma N_F {p_P^2r_h^3\over \hbar^3c}~,
\label{DOS}
\end{equation}
where
$N_F$ is the number of fermionic species. $\gamma$ is dimensionless
constant of order unity. In our oversimplified model it is $\gamma={4\over
35\pi}$ \cite{Volovik}.

The equation of state in the interior region is
$p=\rho \propto T^2$. Incidentally, this coincides with equation of state
of the perfect fluid inside the horizon required to obtain the
Bekenstein-Hawking entropy (see Refs. \cite{ZurekPage,Hooft}
and \cite{MazurMottola}). In Sakharov induced gravity \cite{Sakharov}
the Planck momentum and the gravitational constant are related as
$N_Fp_P^2 \sim \hbar c^3/G$. Actually this means that the microscopic
parameters of the system, the fermion number $N_F$ and the Planck
momentum $p_P$, are combined to form the phenomenological
parameter of the effective theory -- the gravitational contant $G$.  If
one assumes, that only the thermal fermions are gravitating, then one
obtains  $M\sim\int dV\rho\sim T^2 M^3G^2$. This gives estimation
for the temperature and entropy of black hole, $T\sim 1/(GM)$
and $S\sim GM^2$, which is in correspondence with the
Hawking-Bekenstein entropy and the Hawking temperature. Here only the
phenomenological parameters $G$ and $c$ enter, while the microscopic
parameters $N_F$ and $p_P$ drop out. This is in agreement with the
observation made by Jacobson \cite{JacobsonInduced} that the black hole
entropy and the gravitational constant are renormalized so that the
relation between them is conserved. All this means that the statistical
properties of the black hole can be produced by the Fermi liquid in the
interior of the black hole.

\section{Exact energy levels}
\label{ExactLevels}

Another problem, which can be investigated using our scheme, is
concerned with fermion zero modes. Are there fermionic modes which have
exactly zero energy in exact quantum mechanical problem? If yes, this
would justify the conjectures that the black hole has a nonzero entropy
even at $T=0$, and also that the area of the black hole is the quantized
quantity \cite{BekensteinQuantum,Kastrup,Mukhanov}. For this reason we
now proceed to the solution of the eigenvalue equations  (\ref{toka1.44})
and  (\ref{toka1.45}).

\begin{figure}
  \centerline{\includegraphics[width=1.0\linewidth]{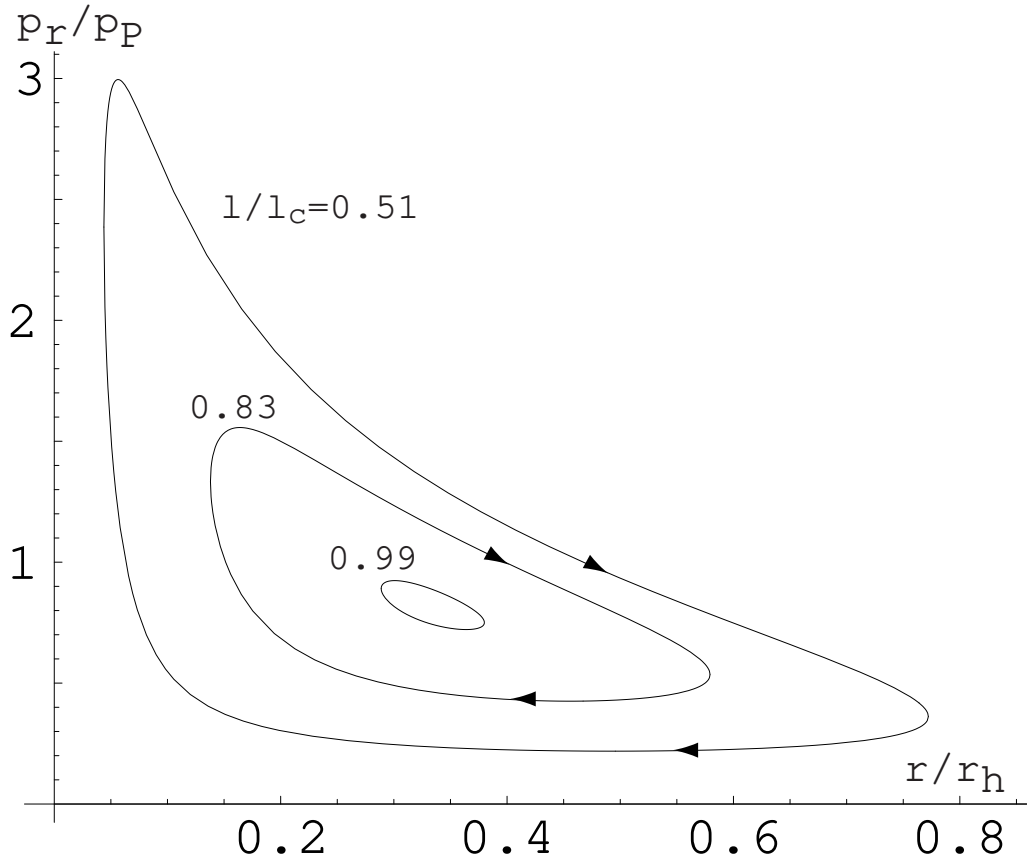}}
  \caption{Closed trajectories of the radial motion inside the black
hole at zero energy $\tilde E=0$ for different values of the angular
momentum $l$. }
  \label{trajectoriesFig}
\end{figure}

It is impossible to solve these equations analytically, but one can choose
the region of parameters, where they can be solved using the perturbation
theory exapnsion in small parameter $1/p_0$. To find this
region, let us consider the quasiclassical trajectories of the radial
motion
$p_r(r)$ at
$\tilde E=0$ and for different
$l$, the Eq.(\ref{SuperluminalTrajectoryZeroE}).
These trajectories are shown in Fig. \ref{trajectoriesFig} (we used
$p_0=10000$). If $l$ is small compared to $p_0$, these
trajectories are highly asymmetric: ingoing and outgoing particles
experience essentially different motion. The conventional relativistic
particles which have small momentum compared with the Planck momentum
$p_P$ can move only towards the singularity. However, when they reach the
large momentum the nonlinear dispersion allows them to move away from the
singularity. As a result the trajectories of particles become closed. This
asymmetry reflects the violation of the time reversal symmetry by the
horizon.

However, when
$l$ increases the trajectories become more and more symmetric. Near the
maximal value
\begin{eqnarray} \label{yh1b}
&&l^{(c)}=3^{-3/2}p_0 \end{eqnarray}
 they become perfectly elliptic and increasingly more concentrated in the
vicinity of the centerpoint
\begin{eqnarray}
\label{yh1a}
&&r^{(c)}={1\over 3} \\ \label{yh1}
&&p^{(c)}=\pm \sqrt{{2\over 3}} p_0~.  \end{eqnarray}
This means that in vicinity of $r^{(c)}$ and $p^{(c)}$ the Hamiltonian
describing the radial motion becomes that of oscillator.
Thus we can expand the equations in the vicinity of $p^{(c)}$ and
$r^{(c)}$ using the small parameter $1/p_0$
\begin{eqnarray}
\nonumber&& r=r^{(c)}+x  \\
&& p_r=p^{(c)}-i\partial_x
~.
\end{eqnarray}
It can be seen that the regions where $x$ and $\partial_x$ are
concentrated
\begin{eqnarray}
x\sim {1\over  \sqrt{p_0}} \ll r^{(c)} \ \ \ \ , \ \partial_x \sim
\sqrt{p_0}\ll |p^{(c)}|\ \ \ \ ,  \end{eqnarray}
become really small compared with  $r^{(c)}$ and $p^{(c)}$ when $p_0$
increases.  As a result,
after lengthy but straightforward expansion of Eq.(\ref{toka1.44}) near
the point with positive $p^{(c)}>0$ one obtains keeping the terms of order
unity the following effective oscillator Hamiltonian:
\begin{eqnarray}
 \nonumber &&H_{eff}= -3\sqrt{{3\over 2}} \delta l +{13
p_0\over 2\sqrt{2}}x^2 +{2\sqrt{2}\over 3p_0 } p^2+{5\over 2\sqrt{3}  }
\left( xp+px \right) +{3\sqrt{3}\over 4}~, \\
\end{eqnarray}
where
\begin{eqnarray}
\delta l \equiv l^{(c)}-(l+1)
\end{eqnarray}
Diagonalization gives the following energy spectrum:
\begin{eqnarray}
\label{kahdes113}
\tilde E_1=-3\sqrt{{3}\over 2}\delta l +
 3n_r +{3\over 2} +{3\sqrt{3}\over 4}~
\end{eqnarray}
where $n_r =0,1,\ldots $ is the radial quantum number.
Accordingly, the expansion near the point with negative $p^{(c)}<0$, and
also the same procedure for the Eq.(\ref{toka1.45}), give the other three
sets of the energy levels:
\begin{eqnarray}
\label{kahdes19}
\tilde E_2=3\sqrt{{3}\over 2}\delta l -
3n_r -{3\over 2} +{3\sqrt{3}\over 4}~,
\end{eqnarray}
 \begin{eqnarray}
 &&
\tilde E_3=-3\sqrt{{3}\over 2}\delta l + 3n_r +{3\over
2} -{3\sqrt{3}\over 4} =-\tilde E_2\end{eqnarray}
and
\begin{eqnarray}
   &&\tilde E_4=3\sqrt{{3}\over 2}\delta l - 3n_r
-{3\over 2} -{3\sqrt{3}\over 4}=-\tilde E_1~.
\end{eqnarray}
Finally, in dimensionful units we have the following discrete levels of
fermions in the vicinity of the Fermi surface:
\begin{eqnarray}
 &&\tilde E(J,n_r) =\pm {\hbar c\over r_h}\left({1\over
\sqrt{2}}{p_Pr_h\over\hbar}- 3\sqrt{{3}\over 2}\left(J+{1\over 2}\right)
- 3n_r -{3\over 2} \pm {3\sqrt{3}\over 4}\right).
\label{DiscreteEnergySpectrum}
\end{eqnarray}
Here all 4 signs must be taken into account. This equation is valid for
$J$ smaller than but close to the maximal value $J^{(c)}=p_pr_h/3\sqrt{3}
\hbar$ at which the states with zero energy can still exist.

The Eq.(\ref{DiscreteEnergySpectrum}) allows us to make conclusion on the
existence of the true fermion zero modes in the presence of black hole.
For general values of $p_Pr_h$, and thus for the general values of the
black hole area $A=4\pi r_h^2$, there are no states with exactly zero
energy. One can find the eigen state with zero energy for some special
values of $A$.  However, because of the incommensurability between radial
and orbital quantum numbers, the degeneracy of the $\tilde E=0$ levels is
small, so that the fermion zero modes cannot produce the entropy at
$T=0$ which is proportional to the area of the horizon.  Accordingly,
there are no microscopic reasons for the quantization of
the area of the horizon.

There are no topological arguments which ensure
the existence the exact fermion zero modes. On the other hand the
momentum-space topology prescribes the existence of the fermion modes
with zero energy on the semiclasical level. These modes form the
surface in the momentum space -- the Fermi surface -- in Fig.
\ref{FermiSurfaceFig}. The existence of the Fermi surface is  the robust
property of the fermionic vacuum and the Fermi surface will survive when
the back reaction will be introduced (of course, if the horizon
survives). It is the Fermi liquid whose thermal states
give rise to the entropy proportional to area, as was discussed in the
previous section.

\section{Conclusion}

In derivation of the fermionic microstates responsible for the
statistical mechanics of black hole we used an analogy between the quantum
liquids and the quantum vacuum -- the ether. We know that in superfluids,
there are two preferred reference frames. One of them is  an
"absolute" spacetime
$(x,t)$ of the laboratory frame, which can be Galilean as well as
Minkowski with $c$ being the real speed of light. In
the effective gravity in quantum liquids, experienced by the low-energy
excitations, the effective `acoustic' metric
$g_{\mu\nu}^{acoustc}$ appears as a function of this "absolute" spacetime
$(x,t)$.  Another preferred reference frame is the local frame, where the
metric is Minkowski in acoustic sense, i.e. with $c$ being the maximum
attainable speed for the low energy quasiparticles. This is the frame
which is comoving with the superfluid condensate. In this frame the
energy spectrum does not depend on the velocity ${\bf v}$ of the
condensate and has a form of Eq.(\ref{EnergySpectrumNonlinear}). This
means that it is in this frame that the Planck energy physice is properly
introduced: if the energy becomes big in the superfluid comoving the
acoustic Lorentz symmetry is violated.

As for the quantum vacuum, the attainable energies are still
so low that we cannot resolve what are the preferred reference frames, if
any. In particular, we cannot say in which reference frame the Planck
energy physics must be introduced, and whether there is an absolute
spacetime. The magnifying glass of the event horizon can serve as
possible source of the  spotting of these reference frames.

In our low-energy corner the Einstein action is covariant: it does not
depend on the choice of the reference frame. That is why the Einstein
equations can be solved in any coordinate system. However, in the
presence of a horizon or ergoregion some of the solutions are not
determined in the whole spacetime of the quantum vacuum. In these cases
the discrimination between different solutions arises and one must choose
between them. In quantum liquids the choice is natural, because from
the very beginning the absolute coordinates are used. But in general
relativity the ambiguity in the presence of a horizon imposes the problem
of the proper choice of the solution. This problem cannot be solved
within the effective theory, while the fundamental "microscopic"
background is still not known, and one can only guess what is the proper
solution of Einstein equations, using which the vacuum state can be
constructed.

It is clear that Schwarzschild solution is not the proper choice,
in particular because the whole spacetime is not covered by Schwarzschild
coordinates. According to quantum liquid analogy, the Painleve-Gullstrand
metric with the frame dragging inward can be the reasonable choice. Its
analog can be really reproduced (at least in principle) in quantum
liquids. The analogy also suggests that the Painleve-Gullstrand spacetime
can be considered as an absolute, in which the true vacuum must be
determined. On the other hand, the local frame of the free-falling
observer can be considered as the analog of the superfluid comoving
frame, in which the Planck energy physics must be introduced. Let us warn
again that from the point of view of the effective theory alone such
choice cannot be justified.

If in addition the Planck physics is superluminal, as is also suggested
by the quantum liquid analogy, the stable quantum vacuum can be
constructed even in the presence of a horizon. We argue that the main
property of such quantum vacuum, which distinguishes it from the original
vacuum of the Standard Model, is the existence of the Fermi surface inside
the horizon. The statistical mechanics of the Fermi liquid formed inside
the horizon is responsible for the thermodynamics of the black hole.

{\bf Acknowledgements.}

GEV thanks Jan Czerniawski and Pawel Mazur for fruitful discussions.
This work was supported by ESF COSLAB Programme. The work of GEV was
supported in part by the Russian Foundations for Fundamental Research.

\end{document}